\documentclass[]{aa} 

\usepackage{graphicx}
\usepackage{txfonts}
\usepackage{natbib}
\usepackage{longtable}
\usepackage{lscape}
\usepackage{color}

\providecommand{\sorthelp}[1]{}

\begin{document}
 
\title{ {DIES: Parallel dust radiative transfer program with the
immediate re-emission method} }

\author{Mika Juvela\inst{1}
}

\institute{
Department of Physics, P.O.Box 64, FI-00014, University of Helsinki,
Finland, {\em mika.juvela@helsinki.fi}
}

\authorrunning{M. Juvela et al.}

\date{Received September 15, 1996; accepted March 16, 1997}

\abstract { 
Radiative transfer (RT) modelling is a necessary tool in the interpretation of
observations of the thermal emission of interstellar dust. It is also often
part of multi-physics modelling. In this context, the efficiency of radiative
transfer calculations is important, even for one-dimensional models.
} 
{
We investigate the use of the so-called immediate re-emission (IRE) method for
fast calculation of one-dimensional spherical cloud models. We wish to
determine whether weighting methods similar to those used in traditional
Monte Carlo simulations can speed up the estimation of  dust temperature.
}
{
We present the program DIES, a parallel implementation of the IRE method,
which makes it possible to do the calculations also on graphics processing
units (GPUs). We tested the program with externally and internally heated
cloud models, and examined the potential improvements from the use of different
weighted sampling schemes. }
{
The execution times of the program compare favourably with previous programs,
especially when run on GPUs. On the other hand, weighting schemes produce
only limited improvements. In the case of an internal radiation source, the
basic IRE method samples the re-emission well, while traditional Monte Carlo
requires the use of spatial importance sampling. Some noise reduction could be
achieved for externally heated models by weighting the initial photon
directions. Only in optically very thin models does weighting ---such as the
proposed method of forced first interaction--- result in noise reduction by a
factor of several.
}
{ 
The IRE method performs well for both internally and externally heated models,
typically without the need for any additional weighting schemes. With run
times of the order of one second for our test models, the DIES program is
suitable even for larger parameter studies.
}


\maketitle

\section{Introduction} \label{sect:intro}

The first goal of dust radiative transfer (RT) modelling is to determine the
temperatures of the dust grains. RT calculations provide information about the
radiation field intensity at each position or cell, which corresponds to the
spatial discretisation of the model. If dust grains are sufficiently large (have
sufficient heat capacity), they can be described using an equilibrium
temperature, which results from the balance of absorbed and emitted energies
and depends on the assumed optical properties of the grains
\citep{Li2001,Draine2003ARAA}. In the conditions of interstellar clouds, the
energy is absorbed mainly at wavelengths from ultraviolet (UV) to
near-infrared (NIR) or even mid-infrared (MIR), where the optical depths can
be high and scattering is important for the transfer of energy. In contrast,
dust re-emits the energy at far-infrared (FIR) wavelengths where the objects
are often optically thin. Thus, the calculations need to consider a wide range
of wavelengths.

The dust RT problem is typically solved with the Monte Carlo
method, by simulating the emission of photons and by following their travel
until they are absorbed or leave the model volume
\citep{TRUST-I,Steinacker2013}. For practical reasons, the large number of
real photons is divided into a smaller number of photon packages that can be
simulated in a computer.  The calculations can be divided into iterations,
where the radiation field is estimated using up-to-date information on the
radiation sources, including the emission from the dust with its current
estimated temperatures. When the simulation step has been finished, the
knowledge of the radiation field is used to update the dust temperatures in
every cell. If the dust re-emission is a significant component of the
radiation field, the process needs to be repeated until the temperatures
converge \citep{Juvela2005}. In the following, we refer to this approach as the
traditional Monte Carlo (TMC) method.

The run times are affected especially by the high optical depths at short
wavelengths. Frequent scatterings and the potentially very short free paths
mean that the information of the radiation field propagates only slowly
through the medium. The time needed to follow a single photon package is a
steep function of the optical depth. Furthermore, if hot dust (with
significant re-emission) fills a small part of the volume, the simulation of
re-emission can become very inefficient. If photon packages are generated at
random locations, few of those will correspond to any significant emission.
This can happen even in spherically symmetric models, when a central radiation
source is surrounded by optically thick but geometrically thin layers of hot
dust \citep{Juvela2005}. The sampling of the radiation field can also become poor
in the innermost cells in externally heated models, because few of the
photon packages initiated at the cloud surface are likely to hit the smallest
cells in the centre.

Some of the problems listed above can be alleviated by weighted sampling, where we adjust some of the probability distributions in the Monte
Carlo simulation and correspondingly adjust the weight of the photon packages
(i.e. the number of real photons they contain). For example, external
radiation could be simulated using photon packages that are directed
preferentially towards the model centre. This improves the accuracy of the
radiation field estimates in the small inner cells. \citet{Juvela2005}
examined such weighting schemes in the context of TMC and one-dimensional
models. Weighting was applied to the initial positions and directions of the
photon packages, as well as the directions of scattered photons. As a special
case, it was demonstrated that weighting can also be used to overcome the
poor propagation of individual photon packages, up to arbitrarily high optical
depths (effectively recovering the diffusion approximation). 

One attribute of Monte Carlo simulations that has been examined by several papers
is the free path, the distance that a photon package travels before
interacting with the medium. In a Monte Carlo program, the random free paths
can be calculated based on the optical depth for absorption or optical depth
for scattering. \citet{Krieger2023} argued in favour of the latter approach
when attempting to improve the estimates for the flux that penetrates through high optical
depths. The free path is just another probability distribution and can be
modified in even more complex ways \citep{BaesGordon2016,CampsBaes2018}.
However, when sampling is improved in one part of the model, it usually deteriorates in other parts. Thus, the goal of weighted sampling could be either to
minimise the maximum errors globally or to minimise the errors for a specific
purpose, such as to better estimate the scattered light from a specific model
region \citep{BaesGordon2016, KriegerWolf2024}. In many TMC programs, the
energy absorption is calculated continuously, with updates made to every cell
along the photon path between discrete scattering events
\citep{Lucy1999,Juvela2005}. This ensures lower noise for the temperature
estimates, especially in regions of low optical depth.

\citet{BjorkmanWood2001} presented a different version of the Monte Carlo
scheme, where absorptions are treated as discrete events. After each
interaction, the temperature of the absorbing cell is updated and the absorbed
energy is immediately reradiated. The new photon package retains the same
energy but is given a new random frequency that reflects the probability
distributions associated with the temperature change. Because of the discrete
locations of the absorption events, the method may suffer from poorer sampling
of regions of low optical depth. On the other hand, as the simulation follows
the actual flow of energy, the dust emission is automatically better sampled
in regions where most of the re-emission takes place (i.e. where dust is hot).
In the following, we refer to the \citet{BjorkmanWood2001} immediate
re-emission method as the IRE method. \citet{Juvela2005} made some comparisons
between the TMC and IRE calculations in the case of spherically symmetric and
mostly optically thick models. The methods were found to exhibit a comparable
performance, although TMC included some non-standard improvements. These included
the use of weighted sampling and the so-called reference field method (to
decrease noise) and accelerated Monte Carlo (to speed up temperature
convergence over iterations). This suggests that the IRE method can be a good
way to estimate dust temperatures, at least if stochastic heating can be
ignored. Since the method involves a temperature update after each absorption
event, it is not clear whether or how IRE could be implemented efficiently in
the case of stochastic heating, where the temperature updates are inherently
much more expensive \citep{Desert1986, DraineLi2001, Camps2015}. IRE
could nevertheless be a good option to be combined for example with the modelling of
chemical cloud evolution, where fast RT is needed and
knowledge of equilibrium dust temperatures may be sufficient
\citep[][]{Sipila2020, Chen2022COM}.

In this paper, we present a new RT program, `dust interactions of emission and
scattering' (DIES), which implements the IRE algorithm for parallel
computations on both multi-core central processing units (CPUs) and graphics
processing units (GPUs)\footnote{Available at https://github.com/mjuvela/ISM}. We use a series of test problems to investigate the
performance of the IRE method and to study whether or not IRE can be enhanced
with the use of weighted sampling, as in the case of the TMC methods.

The paper is organised as follows. The IRE method and the weighting schemes
are described in Sect.~\ref{sect:methods}. A series of test cloud models are
presented and potential improvements from weighted sampling are quantified in
Sect.~\ref{sect:results}. The results are discussed in
Sect.~\ref{sect:discussion} before listing the main conclusions in
Sect.~\ref{sect:conclusions}.

\section{Methods}  \label{sect:methods}

\subsection{Basic IRE method} \label{sect:methods_basic}

The implementation of the IRE method in DIES follows the description in
\citet{BjorkmanWood2001}, in the context of models discretised into spherical
cells or shells. There are two potential sources of radiation, an external
isotropic radiation field and a point source at the centre of the model. 

In the basic version, the total emission per unit time from each of these
sources is divided between $N$ photon packages of equal energy. We keep track of the free paths for scattering and for absorptions 
separately. Both are
initialised with random values of optical depth,
\begin{equation}
\tau_{\rm S}=-\ln u_1, \quad  \tau_{\rm A}=-\ln u_2,
\label{eq:taus}
\end{equation}
where $u_1$ and $u_2$ are independent uniform random numbers, $0 \le u \le 1$.
The initial frequency of a photon package is generated from the probability
distribution dictated by the source spectrum. We use precalculated lookup
tables so that a uniform random number can be quickly mapped to a random
frequency within the source spectrum.

Thanks to the model symmetry, the photon package entering from the background
can be always initialised at the position $(x,y,z)=(0,0,-R_0)$, where $R_0$ is
the outer radius of the model cloud. The angle $\theta$ with respect to the
normal of the surface (the normal vector pointing inwards) is generated as
\begin{equation}
  \mu = \cos \theta = \sqrt{u_3},
\end{equation}
with another uniform random number $u_3$. The resulting initial direction is
\begin{equation}
(d_{\rm x}, d_{\rm y}, d_{\rm z}) = (\cos \phi \sin \theta, \sin \phi \sin
\theta, \cos \theta).
\end{equation}
The rotation angle $\phi$ is in principle uniformly distributed over $0 \le
\phi < 2 \pi$, but the value is irrelevant given the model symmetry. For
photons emitted by the central source, the procedure is the same, except that
$\mu = \cos \theta$ now has a uniform distribution in the range $-1 \le \cos
\theta \le 1$.

A photon package is followed in its original direction one cell at a time, and
the remaining free paths are correspondingly updated:
\begin{equation}
\tau_{\rm A}  = \tau_{\rm A} - \kappa_{\rm A, \nu} \rho l  \quad {\rm and} \quad
\tau_{\rm S}  = \tau_{\rm S} - \kappa_{\rm S, \nu} \rho l,
\label{eq:taus2}
\end{equation}
where $l$ is the step length, $\kappa_{\rm A, \nu}$ and $\kappa_{\rm A,
\nu}$ are the dust absorption and scattering coefficients (per mass for the
current frequency), and $\rho$ is the density of the cell.

If $\tau_{\rm S}$ becomes negative before the end of the step (and before
$\tau_{\rm A}$ becomes zero), the step is truncated to the position where
$\tau_{\rm S}$ reaches zero. The photon package is scattered in a new
direction according to the dust scattering function $\phi_{\nu}$, and both
$\tau_{\rm A}$ and $\tau_{\rm S}$ are assigned new random values from
Eq.~(\ref{eq:taus}).

If $\tau_{\rm A}$ becomes negative before the end of the step (and before
$\tau_{\rm S}$ becomes zero), the step is truncated to the position where
$\tau_{\rm A}$ reached zero. The energy of the photon package is absorbed into
the current cell, and the temperature of the cell is updated. The same amount
of energy is then re-emitted by creating a new photon
package in the same position with a random direction (uniform probability over the full solid
angle). The frequency for the photon package is generated from the probability
distribution that corresponds to the difference between the emission spectra
at the new and old temperatures. Both the new dust temperature and the
resulting new frequency are obtained using precalculated lookup tables.
Finally, both $\tau_{\rm A}$ and $\tau_{\rm S}$ are again given new values
according to Eq.~(\ref{eq:taus}). The calculation ends when all photon
packages from the background and from the point source have been followed
until they exit the model volume.

\begin{table}
\caption{Weighted probability distributions and their parameters}
\begin{tabular}{ll}
\hline
\hline
Description  &  Symbol \\
\hline
 Direction of background photons &   $k_{bg}$     \\
 Direction of re-emitted photons &   $g_{\rm E}$  \\
 Direction of scattered photons  &   $g_{\rm S}$  \\
 Free path for absorption        &   $k_{\rm A}$  \\
 Free path for scattering        &   $k_{\rm S}$  \\
\hline 
\end{tabular}
\label{table:parameters}
\end{table}

\subsection{Weighted sampling} \label{sect:methods:weighted}

We tested weighted sampling for five probability distributions that appear in
the Monte Carlo simulation. These can be divided to the weighting of
directional and distance probability distributions. The options (cf.
Table~\ref{table:parameters}) are described below.

The weighting means that the original probability distribution is replaced
with a new distribution from which the direction or distance is generated.
This change is compensated by adjusting the relative weight of the photon
package (i.e. changing the number of true photons that it contains). The
weighting factor is the ratio between the original and the modified
probabilities that are estimated using the generated value of the variable (i.e. a
specific direction or value of a free path). Therefore, alternative
distribution must be positive for all valid variable values, and the probability
ratio should have only a moderate range of values below and above one  in order
to avoid
noise spikes caused by occasionally very large weight factors. The
goal of the weighting is to increase the sampling (decrease the noise) in some
parts of the parameter space, especially those where the noise would otherwise
be the largest.

\subsubsection{Directional weighting}

The probability distributions for the direction of photon packages appear in
the initialisation of background radiation, the direction of re-emitted
photons, and the direction of scattered photons. Because of the spherical
symmetry, weighting is not applied to the photon packages from the central
source, where the distribution is always isotropic.

When a spherical model is illuminated by external radiation, the smaller
projected size of the innermost cells results in higher temperature
errors  there. The problem may be alleviated by directing the photon packages
preferentially towards the model centre. We implemented this by replacing the
default direction distribution $\mu = \cos \theta \sim \sqrt{u}$ with $\mu \sim
u^{k_{\rm bg}}$. Each photon package is correspondingly weighted by a factor 
\begin{equation}
W(\mu, k_{\rm bg}) = 2 \, k_{\rm bg} \, \mu^{2-1/k_{\rm bg}},
\end{equation}
using here the current realisation of the $\mu$ value. Thus, the value $k_{\rm
bg}=0.5$ corresponds to the original, unweighted case, and values $k_{\rm
bg}<0.5$ result in more photon packages being directed towards the central
regions of the model.

We can similarly modify the direction of the photon packages after each
interaction with the medium. In the case of re-emission, we replace the
original isotropic distribution of directions with a Henyey-Greenstein
function \citep{Henyey1941},
\begin{equation}
p(\mu) =  \frac{1}{2} \frac{1-g_{\rm A}^2}{({1+g_{\rm A}^2-2 g_{\rm A} \mu})^{3/2}}.
\end{equation}
In the case of external heating, the direction $\mu= \cos \theta$ is measured
with respect to the direction towards the model centre, and in the case of a
central point source with respect to the opposite direction. Thus, with
positive values of the asymmetry parameter $g_{\rm A}= \langle \cos \theta
\rangle$, photon packages are sent preferentially in the direction away from
the original source. In particular, in  this way, photon packages from the external
radiation field should have a greater chance of reaching the model
centre.

The procedure is similar for scattered photons, except that the original
distribution of directions is not isotropic but depends on the direction of
the photon package before the scattering event and the dust scattering
function. We use Henyey-Greenstein functions for both of these. If the
scattering angle is $\mu = \cos \theta$ relative to the original direction of
propagation and $\mu' = \cos \theta'$ is relative to the radial reference
direction, the weight is now the ratio of the two Henyey-Greenstein probabilities,
\begin{equation}
W(\mu, \mu^{\prime}, g_{\rm S}) = \left( \frac{1-g^2}{ (1+g^2-2 g \mu)^{3/2}} \right) / \left(
\frac{1-g_{\rm S}^2}{ (1+g_{\rm S}^2-2 g_{\rm S} \mu^{\prime})^{3/2}}
\right)^{-1}.
\end{equation}
Here $g$ is the asymmetry parameter of the dust scattering function and 
$g_{\rm S}$ is the parameter chosen for the weighted sampling.

If both the remission and scattering directions are weighted strongly towards
the model centre (and the free paths are short), it is possible for a photon
package to remain trapped within that region. As a safeguard, DIES will
terminate a photon package if its weight has decreased by ten orders of
magnitude. If the trapping is caused by the weighting, the limit will be
reached after some tens of scattering or re-emission events. However, such 
strong weighting will also almost certainly result in higher noise in the
temperature estimates.

\subsubsection{Weighting of free paths}

The remaining weighting options concern the free paths, where the goals depend
on the model optical depths. When optical depths are high, one could penetrate
layers of high optical depth more efficiently by biasing the distribution to
contain a greater number of long free paths. Conversely, if the medium is optically thin,
shorter free paths could mean more frequent absorption events (for photon
packages of smaller weight) and thus lower noise.

The original probability distribution of the free path is $p(\tau)=e^{-\tau}$,
when expressed in units of the optical depth. We implemented the weighting
separately for absorption and scattering events. We use a simple scaling with
a constant $k$ ($k_{\rm A}$ for absorption and $k_{\rm S}$ for scattering).
The modified probability distribution is $p^{\prime}(\tau)=e^{-k \tau}$, meaning
that values $k>1$ lead to shorter average free path, and the relative weight
of a photon package is 
\begin{equation}
W(\tau, k) = \frac{1}{k} e^{\tau(k-1)}.
\end{equation}

Some tests were made with density-dependent factors, where $k$ is a linear
function of $\log n$, based on the local density $n$. The lowest densities
corresponded to the largest values of $k$ (preferring short free paths), and thus
induce more absorption events in low-density regions.

The final option concerning free paths could be called forced first
interaction (FFI), in analogy to the method of forced first scattering that is
often used in calculations of scattered light \citep{Mattila1970, Witt1977}. This
could be useful for models of low optical depth, where a large fraction of the
photon packages would not otherwise interact with the medium at all. The
method requires additional work to first calculate the total optical depth
$\tau_0$ along the original photon direction. The normal cumulative
probability distribution of free paths, $P(\tau)=1-e^{-\tau}$ , is replaced with
the conditional version
\begin{equation}
P(\tau)=\frac{1-e^{-\tau}}{1-e^{-\tau_0}}, \quad P(\tau_0)=1.
\end{equation}
and the original formula of free paths is replaced with
\begin{equation}
\tau = - \ln [ 1- u(1-e^{-\tau_0}] .
\end{equation}
The modified probability distribution is now
\begin{equation}
p(\tau) = \frac{e^{-\tau}}{1-e^{-\tau_0}},
\end{equation}
meaning that the relative weight of a photon package becomes
\begin{equation}
W(\tau) = 1-e^{-\tau_0}.
\end{equation}
Above, the optical depth is taken to stand for the sum of absorption and
scattering. Stronger biasing could also be accomplished by using the
optical depth of absorption only.

\subsection{DIES implementation} \label{sect:implementation}

DIES implements IRE and the above-described weighting methods in a Monte Carlo
program. The program uses OpenCL libraries, which enable parallel execution on
both multi-core CPUs and GPUs. The performance is expected to be good,
especially on GPUs, where modern hardware makes it possible to simulate 
thousands of photon packages in parallel.

Most GPUs have much higher performance for single-precision than for
double-precision floating point operations. Therefore, DIES uses only single
precision. However, initial tests showed that this can lead to problems in
simulations with large numbers of photon packages ($\sim 10^8$ or more,
depending on the model). This happens when the total energy absorbed in a cell
is many orders of magnitude larger than the update provided by a single photon
package. The updates to the absorption counters first suffer from loss of
precision and eventually the updates get rounded to zero. This leads to the
temperatures being underestimated in those parts of the models where the
number of updates is the largest.

In DIES, the problem of floating point precision is avoided by dividing the
calculations into smaller batches. One uses one counter for the absorptions
during all previous batches and one for the current batch. If the latter
contained only one update  at that moment, the sum of the two counters can
still suffer from the above-mentioned rounding problem. However, that is by
definition at the limit of the precision of the 32 bit floating point numbers
(about seven decimal places) and is insignificant when the frequency of a
re-emitted photon package is generated. For a sufficiently large number of batches,
the counter for the current batch never grows so large that it will itself
become affected by rounding errors. Similarly, the sum of the two counters
always retains almost the full precision of a 32 bit float, especially
concerning the final estimates of the temperatures. By using $N_{\rm B}$
batches, the problem of the rounding errors is thus deferred to runs with a
factor of $N_{\rm B}$ larger total number of photon packages. Other solutions
to the rounding problem (e.g. direct use of double precision or the Kahan
summation \citep{Kahan1965} or other similar algorithms) appear to be either
slower or even impossible to implement on GPUs, because of the lack of
synchronisation between GPU threads.

\section{Results} \label{sect:results}

We tested the performance of the DIES program using a series of model clouds
that are heated either externally by the interstellar radiation field
\citep{Black1987} or internally by a central point source emitting 
$T=10000$\,K blackbody radiation of one solar luminosity. The density
distributions were calculated based on the model of critically stable
Bonnor-Ebert spheres \citep{Ebert1955,Bonnor1956} at a gas temperature of
15\,K, discretised into 100 shells of equal thickness. The mass of the spheres
was varied between 0.2\,$M_{\odot}$ and 10\,$M_{\odot}$. The tests were carried
out with the dust model of \citet{Compiegne2011}, including its non-isotropic
scattering functions approximated with Henyey-Greenstein functions. The lower
model masses correspond to higher optical depths and thus to higher
temperature gradients between the model centre and its surface layers. The
visual extinctions to the model centre are $A_{\rm V}=78.5^{\rm m}$ and
$A_{\rm V}=1.6^{\rm m}$ for the 0.2\,$M_{\odot}$ and 10\,$M_{\odot}$ models,
respectively.

\begin{figure}
\begin{center}
\includegraphics[width=9.0cm]{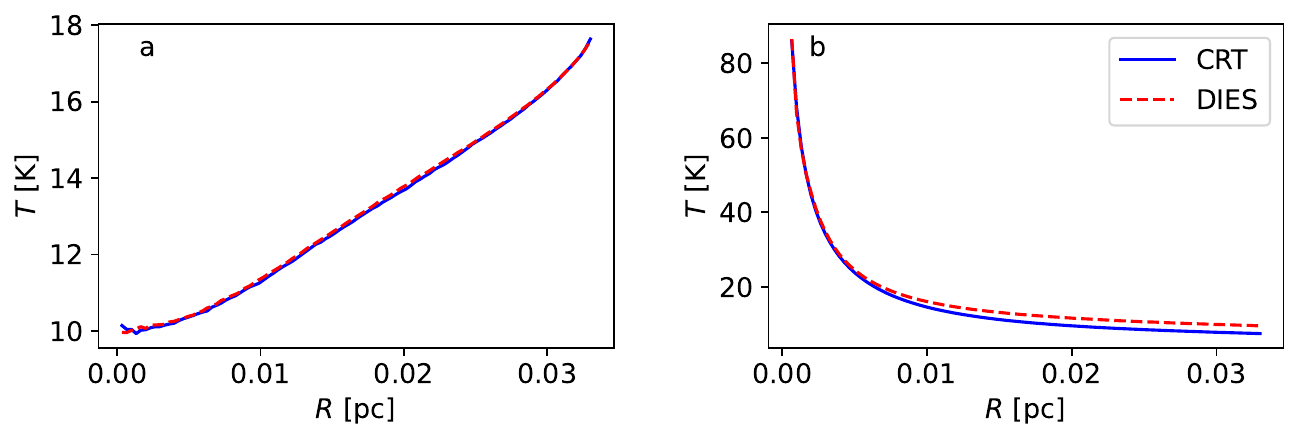}
\end{center}
\caption { 
Examples of temperature distributions in cases of  externally heated (frame a)
and internally heated (frame b) models. The density distribution is that of a $1
M_{\odot}$ critically stable Bonnor-Ebert sphere. The blue and dashed red
lines show the temperatures calculated with the CRT \citep{Juvela2005} and
DIES programs, respectively.
}
\label{fig:T}
\end{figure}

Figure~\ref{fig:T} shows an example of the temperature profiles in the case of
1\,$M_{\odot}$ models (with $A_{\rm V}=15.7^{\rm m}$ to the centre of the
model). The plots also show temperature distributions that were
calculated with the Continuum Radiative Transfer (CRT) program
 for comparison \citep{Juvela2005} (based on TMC rather than the IRE method).

\subsection{Basic tests}  \label{sect:basic}

\begin{figure}
\begin{center}
\includegraphics[width=9.0cm]{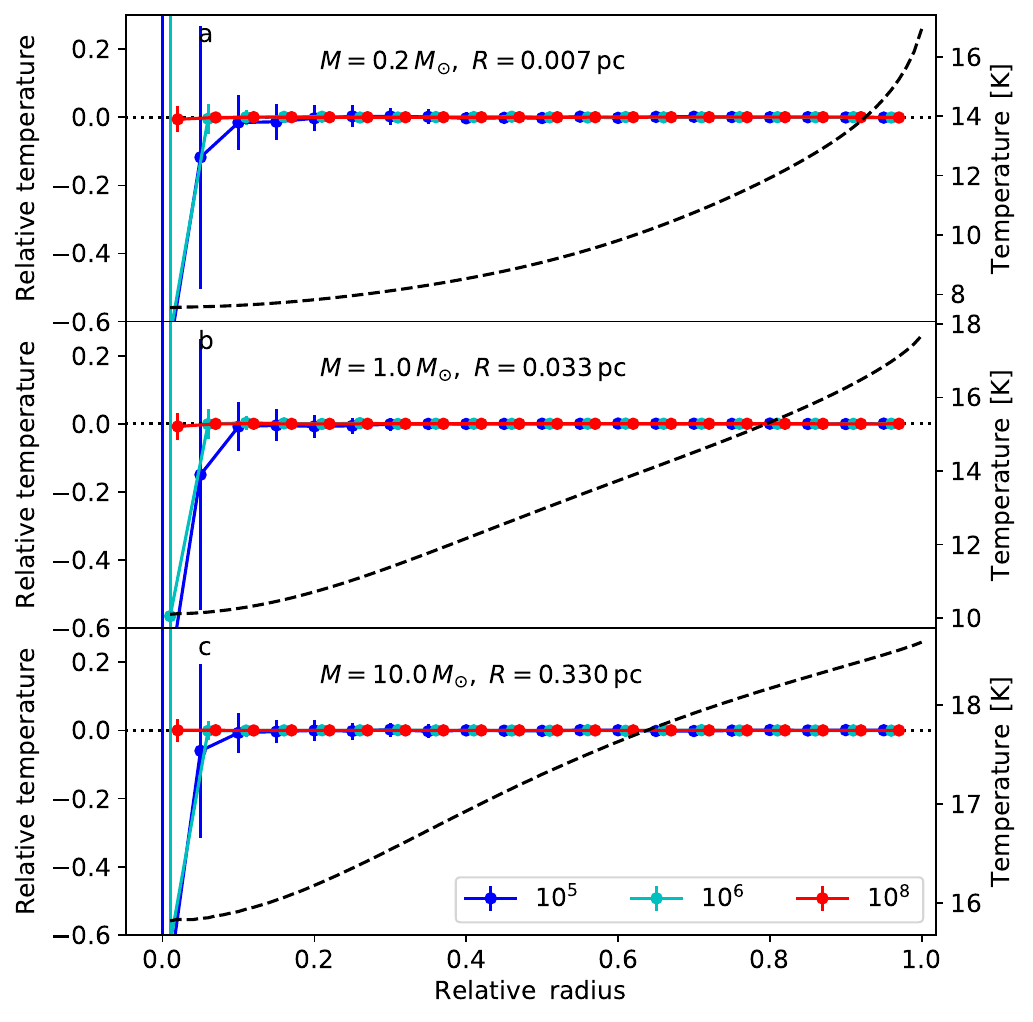}
\end{center}
\caption { 
Accuracy of temperature estimates for externally heated models as a function
radial position. The frames correspond to 0.2, 1.0, and 10\,$M_{\odot}$ mass
models, with the outer radii given in the figure. The dashed black line and
the right axis indicate the dust temperatures calculated with $10^9$ photon
packages. Compared to this reference solution, the blue, cyan, and red symbols
show the results (temperature ratios) for smaller samples of $10^5$, $10^6$,
and $10^8$ photon packages. These are plotted for every fifth cell, starting
with the centre cell, and adding small shift along the x-axis for better
readability. The error bars show the 1$\sigma$ temperature dispersion based
on 100 independent runs.
}
\label{fig:timeit_bg}
\end{figure}

Figure~\ref{fig:timeit_bg} shows the systematic and statistical temperature
errors for externally heated model clouds, based on runs with 10$^5$-10$^9$
photon packages. The temperature errors are below a few percent in the outer
parts, within 10\% of the outer radius. For lower numbers of photon packages,
the temperatures tend to be underestimated within the model centre and the
statistical errors increase rapidly as only a small fraction of the photon
packages reach the innermost cells. However, irrespective of the model mass
and optical depth, 10$^8$ photon packages is sufficient to reach $\sim 1$\%
accuracy for all cells.

The calculations employed $N_{\rm B}=50$ batches. However, even without that
(i.e. for $N_{\rm B}=1$) the effect of rounding errors is visible only in the
surface layers of the $0.2\,M_{\sun}$ model and only when the number of photon
packages is increased to $10^9$.

\begin{figure}
\begin{center}
\includegraphics[width=9.0cm]{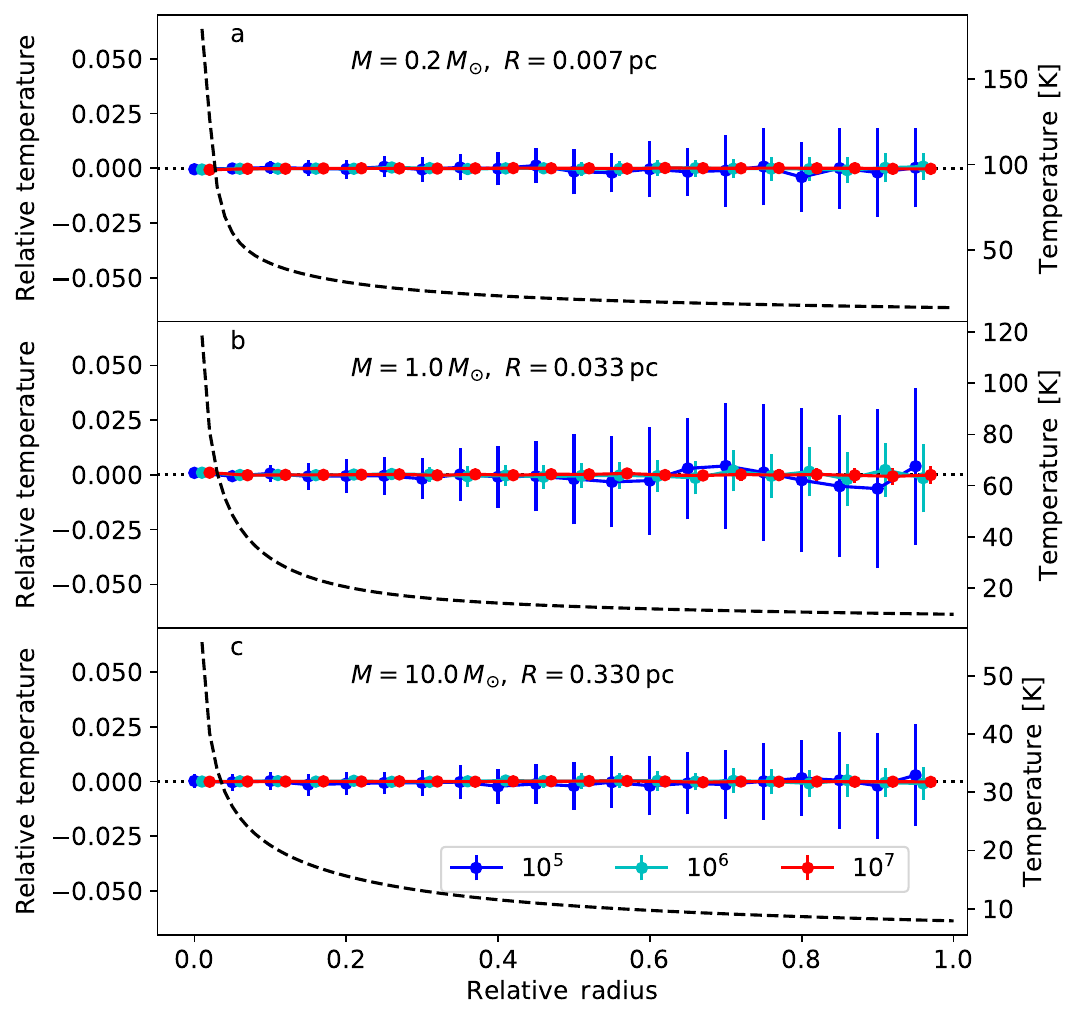}
\end{center}
\caption { 
As in Fig.~\ref{fig:timeit_bg} but for models heated by a central source. The
results are shown for $10^5$, $10^6$, and $10^7$ photon packages, relative to
runs with $10^{8}$ photon packages.
}
\label{fig:timeit_ps}
\end{figure}

Figure~\ref{fig:timeit_ps} shows a similar plot of errors for internally
heated models. As expected, the errors now increase outwards, but the
statistical errors remain below $\sim 3$\%, even with just $10^5$ photon
packages. With $10^8$ photon packages, the rounding errors of $N_{\rm B}=1$
calculations result in clear errors towards the centre of both the
0.2\,$M_{\odot}$ and 1.0\,$M_{\odot}$ models. These disappear already with
$N_{\rm B}=10$, and thus the estimated
central temperatures are also identical from $N=10^5$ to $N=10^8$  in Fig.~\ref{fig:timeit_ps}.

\begin{figure}
\begin{center}
\includegraphics[width=9.0cm]{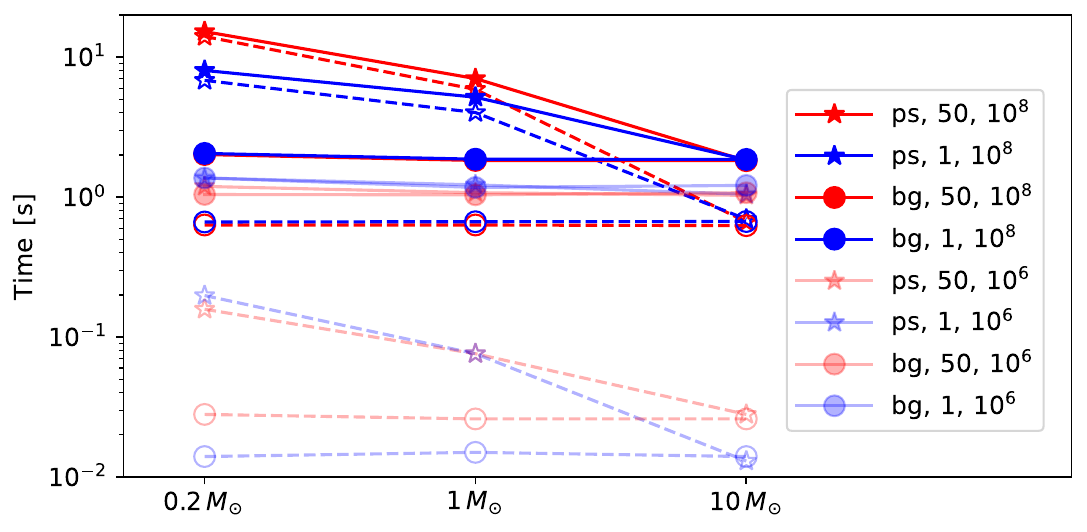}
\end{center}
\caption { 
Run times for $10^6$ (transparent colours) and $10^8$ (solid colours) photon
packages in the case of the three model clouds with external heating (circles)
or internal heating (stars). The solid lines and filled symbols indicate the
total run times (including the Python host script) and the dashed lines and
open symbols indicate the time of the actual simulation (the OpenCL kernel).
}
\label{fig:times}
\end{figure}

Figure~\ref{fig:times} shows a comparison of some execution times in runs with
$10^6$ and $10^8$ photon packages. For external heating, these are dominated
by constant overheads (mainly in the Python host program), and the actual
simulation takes less than one second. 
The overheads include the preparation of the lookup tables (for the
calculation of dust temperatures and for the generation of the frequencies of
the re-emitted photon packages) and the on-the-fly compilation of the OpenCL
kernels. These are independent of both the model size and the number of photon
packages used.
As a result, the hundred-fold increase in the number of photon packages
translates to less than a factor of two increase in the actual run time. In
contrast, for internally heated models and $10^8$ photon packages, most of the
time is spent in the actual simulation. These run times also depend on the
optical depths, with the difference between the $0.2\,M_{\odot}$ and
$10\,M_{\odot}$ models approaching one order of magnitude. The plot also shows
that the overhead from splitting the calculations to 50 batches is at most
a factor of two. These overheads affect mostly internally heated models, as
those run times are dominated by the simulation.

The run times of the CRT and DIES programs are compared shortly in
Appendix~\ref{app}. The tests cover one externally heated and one internally
heated cloud model. 

\subsection{Tests of weighted schemes}  \label{sect:weighted}

\begin{figure}
\begin{center}
\includegraphics[width=9.0cm]{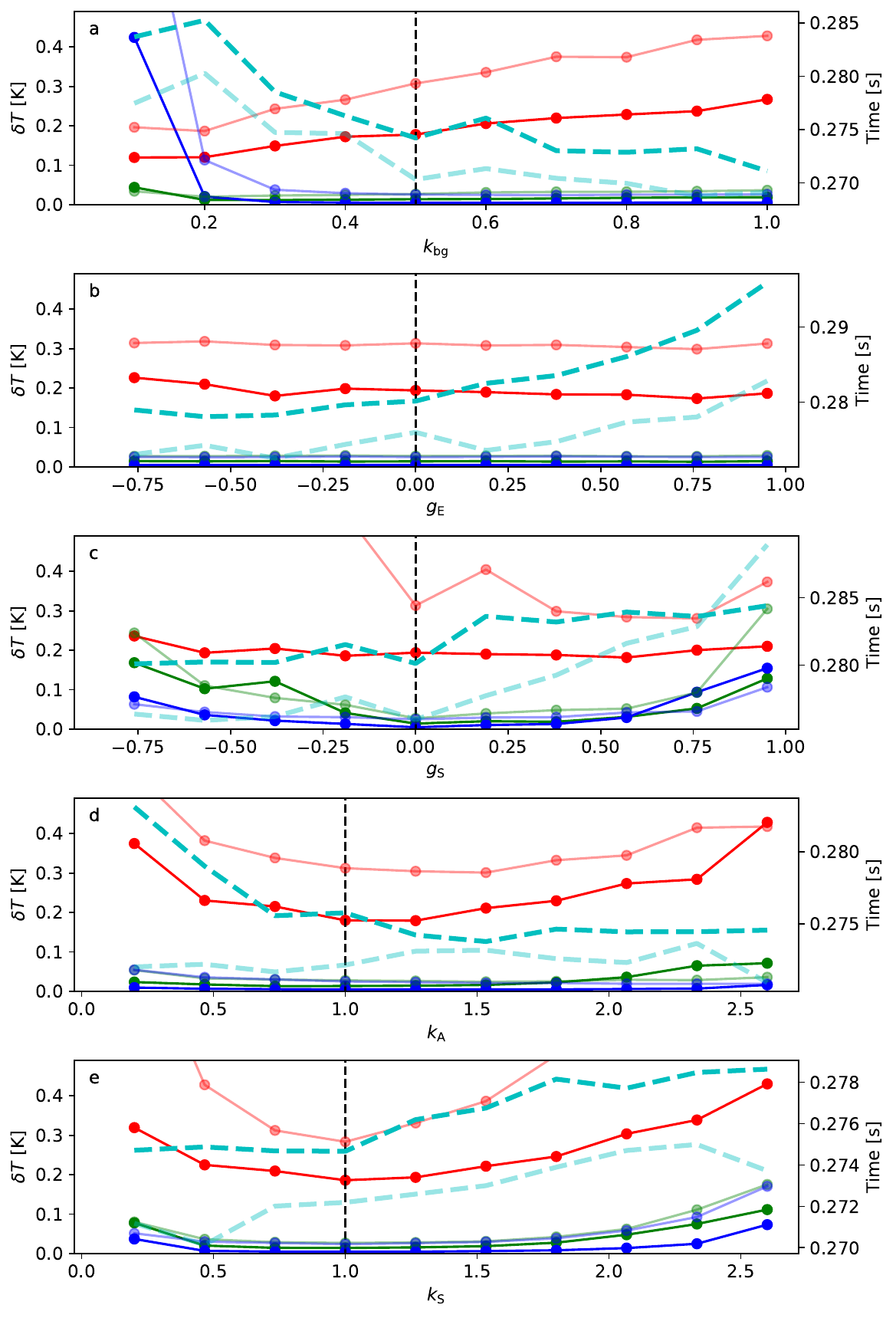}
\end{center}
\caption { 
Relative accuracy of temperature estimates in the case of weighted sampling. Each
frame corresponds to one of the weighting parameters listed in
Table~\ref{table:parameters}. The errors are shown for the surface (blue
curves), the half-radius point (green curves), and the centre of the model.
The vertical dashed lines mark the default solution without weighting. The
dark lines correspond to the 0.2\,$M_{\sun}$ and the light (transparent)
curves to the 10\,$M_{\sun}$ models. The cyan lines and the right hand axis
show the corresponding run times for the simulation step only.
}
\label{fig:plot_error}
\end{figure}

Figure~\ref{fig:plot_error} shows how weighted sampling affects the precision
of the temperature estimates for the externally heated $0.2\,M_{\sun}$ and
$10\,M_{\sun}$ models. The runs used $10^7$ photon packages, and weighting
was applied to only one probability distribution at a time (cf.
Table~\ref{table:parameters}).

Biased sampling of the initial package directions with $k_{\rm bg}<0.5$ does
reduce errors in the model centre. The noise is almost a factor of two lower
for $k_{\rm bg}=0.2$ (Fig.~\ref{fig:plot_error}a). This corresponds to almost
a factor of four in the run times (for a given noise level), because the run
time per photon package remains almost constant. With small values $k_{\rm
bg}<0.2$, the errors increase in the outer model layers and even exceed those
of the innermost shells. The effect $k_{\rm bg}$ is similar for both cloud
models, in spite of their different optical depths. Other options of
directional weighting have little effect on the noise levels. For the
parameter $g_{\rm E}$ this is natural, because dust remains too cold for the
dust re-emission to be of any significance. For the $g_{\rm S}$ parameter, the
errors increase for both large positive and negative parameter values and
especially in the model surface layers.

The weighting of the free path of absorptions with $k_{\rm A}>1$ was expected
to improve the sampling at low densities and thus in the outer parts of the 
$M=0.2\,M_{\sun}$ model and more uniformly in the lower-density
$M=10\,M_{\sun}$ model. However, this is not observed in practice, possibly
because even the $M=10\,M_{\sun}$ model is not optically thin. The default
unweighted method is in these cases nearly optimal.

\begin{figure}
\begin{center}
\includegraphics[width=9.0cm]{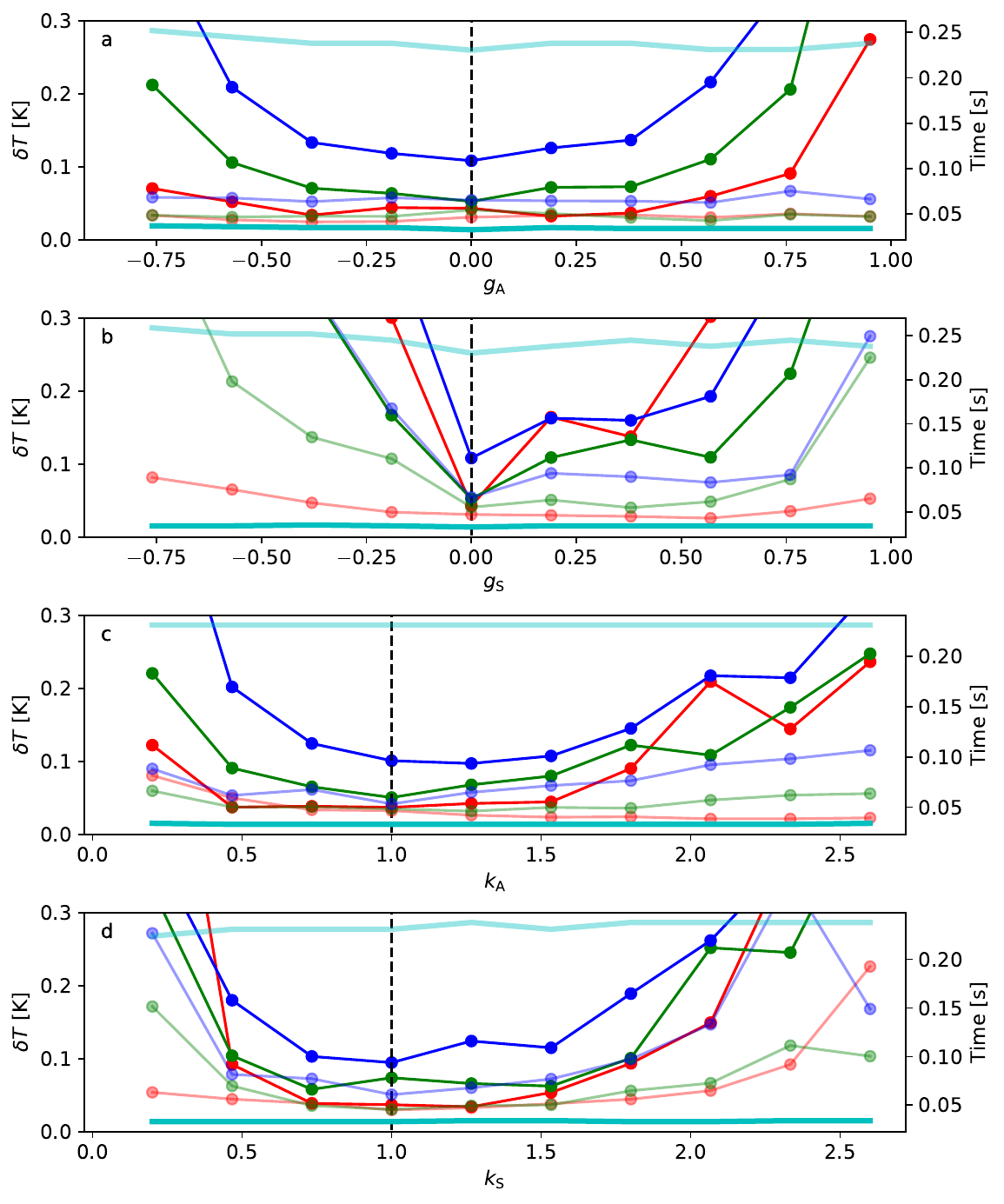}
\end{center}
\caption { 
As in Fig.~\ref{fig:plot_error} but for the internally heated models.
}
\label{fig:plot_error_ps}
\end{figure}

Figure~\ref{fig:plot_error_ps} shows the corresponding results for the
internally heated models. The parameter $k_{\rm bg}$ is omitted because it is
not applicable to internal heating. The unweighted DIES calculations are again
nearly optimal regarding all four weighting schemes. In the plots, the
vertical dashed lines always correspond to the default method without
weighting. 
For most parameters, this is automatically true, and for example when the
directions of the background photon packages are weighted with $k_{\rm bg}=0.5$,
the calculations are identical to the normal, unweighted sampling.
However, in the case of scattering directions, the parameter $g_{\rm S}=0$
(isotropic directions of scattered photon packages) does not correspond to the
normal run, where the scattering directions are anisotropic due to the dust
scattering function. However, for reference, we still include the results from
the normal, unweighted run in Fig.~\ref{fig:plot_error_ps}c by plotting them
at the location $g_{\rm S}=0$.

Overall, most weighting schemes of Table~\ref{table:parameters} are not
useful for the examined models. The one exception is the $k_{\rm bg}$
parameter in the case of externally heated models. The relevance of this
parameter would increase if the size of the innermost cells were even smaller
or if the models were to have a lower optical depth (with fewer scatterings
masking the effect of the initial package directions). On the other hand, the
weighting methods also did not have a clear effect on the run times. The
timing variations are small and affected by random factors, such as the
computer power management. Thus, for example, the apparent trend in
Fig.~\ref{fig:plot_error}d in the case of the 0.2\,$M_{\odot}$ model is not
reproducible.

\begin{figure}
\begin{center}
\includegraphics[width=9.0cm]{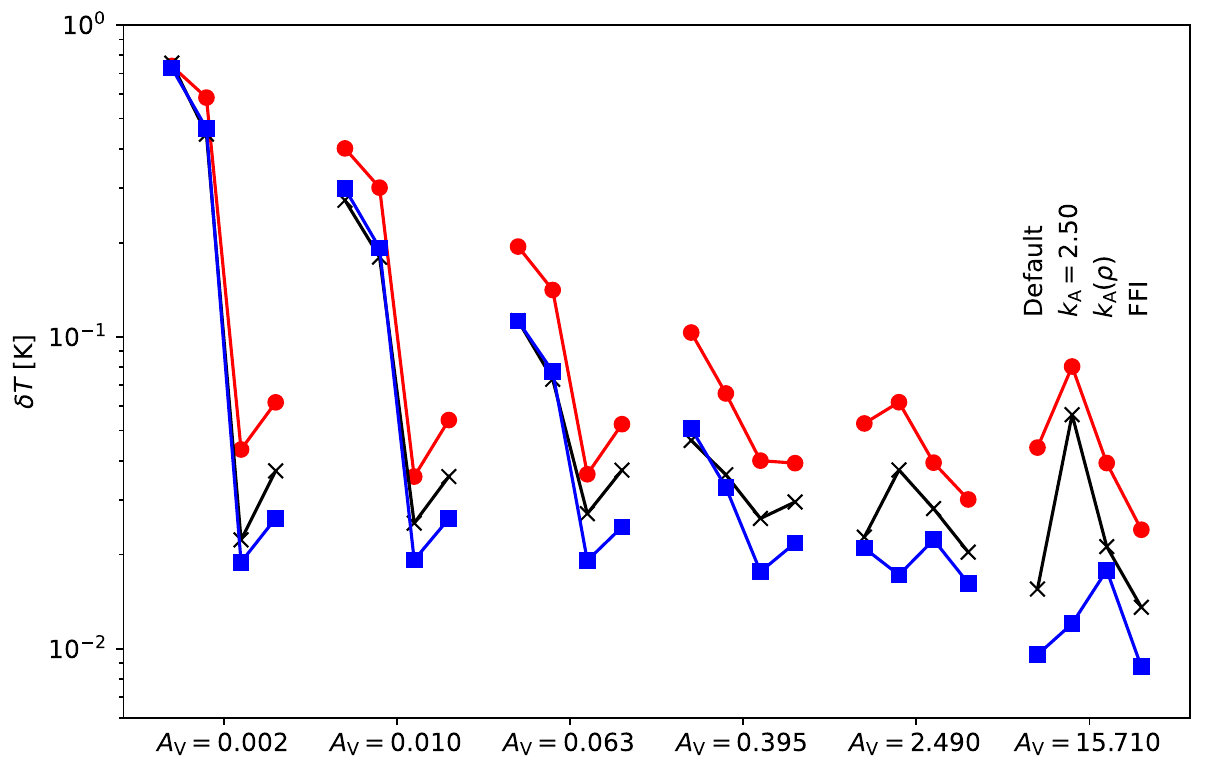}
\end{center}
\caption { 
Test of externally heated models of different optical depth, as indicated by
the $A_{\rm V}$ values on the x axis. The noise of the temperature estimates
is shown for the centre (red symbols, values scaled by 0.1), half radius
(black symbols), and outer surface (blue symbols) of the model clouds. For
each $A_{\rm V}$, results are shown (each group of connected symbols, from
left to right) for the default method, the use of fixed $k_{\rm A}=2$ and
density-dependent $k_{\rm A}(\rho)$ parameters, and for the FFI method.
}
\label{fig:plot_error_special}
\end{figure}

The idea of FFI is somewhat similar to the biasing of the free paths, which
were not useful in the above models. However, if the optical depths are low
enough, these must have a positive effect on the noise levels. We made further
comparisons between calculations with the default method (unweighted), a fixed
value of $k_{\rm A}=2.5$, a density-dependent $k_{\rm A}$, and the FFI method
(Fig.~\ref{fig:plot_error_special}). In the density-dependent case, $k_{\rm
A}(\rho)$ was set to the value of 3.0 at the lowest densities and to 1.0 (i.e.
unmodified free paths) at the highest densities. The parameter was varied
linearly as a function of the density logarithm. We used the
10\,$M_{\odot}$ cloud model in this test, but scaled all density values with a constant
scaling factor to obtain models of different optical depths. For models similar
to or more opaque than the original 10\,$M_{\odot}$ model (i.e. $A_{\rm
V}>1.5^{\rm m}$), all four alternatives result in similar noise
levels (within a factor of two), although the path-length scaling (fixed or
variable $k_{\rm A}$) tends to be poorer than with the default method and FFI
is better than the default method. However, at lower optical depths the default
method performed the worst and the density-dependent biasing and FFI both
reduced the noise significantly, that is, by up to one order of magnitude for the
models with the lowest optical depths. The default method and FFI show almost identical run times per photon package, and so this would imply
a saving of a factor of almost one hundred in the run time needed to reach a given noise
level. The general significance of this is limited by the fact that the
highest speed-ups are seen only in very diffuse models. It is also interesting
to note that the use of a fixed $k_{\rm A}$ value results in far poorer
results than the variable $k_{\rm A}(\rho)$, and this remains true even when
other fixed $k_{\rm A}$ values in the range of $1-3$ are tested.

\section{Discussion} \label{sect:discussion}

The run times of DIES compare favourably for example with the CRT program
\citep{Juvela2005}. The codes use different Monte Carlo schemes: DIES uses the IRE
and CRT the TMC method. The IRE method is sometimes said to contain no
iterations \citep{BjorkmanWood2001}, while the traditional method requires
that the RT simulation and the temperature updates be
iterated until convergence has been reached. 
However, the difference can also be seen more as a reorganisation of the
same of similar calculations.
IRE has the
advantage that the temperatures are always consistent with the current state
of the simulation, although this advantage is offset by the fact that the number of individual
temperature updates is higher, which becomes a problem when this entails large
costs. Thus, it is still not clear how the IRE method could be implemented
efficiently so that also the emission from stochastically heated grains could
be estimated.

The performance difference between CRT and DIES is mostly due to the better
parallelisation of the latter and especially the possibility of using GPUs for
massive parallelisation. The foreseen main use cases of IRE are simple
spherical models with a moderate number of cells and a moderate difference
between the radii of the smallest cells and the model.
Therefore, the tested models  also consist only of 100 shells and have a ratio of
1:100 between the smallest cell and the model radius. Whether illuminated only
by an internal source or only by an external radiation field, runs with $10^8$
photon packages and run times of $\sim$1 second were enough to determine dust
temperatures with an accuracy of $\delta T =0.1$\,K. If combined for example
with chemical modelling (as mentioned in Sect.~\ref{sect:intro}), such
precision is usually sufficient. The errors behaved in a predictable manner,
increasing inwards in the externally heated models and outwards in the
internally heated ones. If a model includes both types of radiation source, the
maximum errors should be even better constrained.

We investigated several possibilities of weighted sampling, but in general did not find these to yield any significant improvement beyond the standard IRE
version.  \citet{Juvela2005} discussed at least two weighting schemes that
were very beneficial in the context of the TMC scheme. The first concerns
externally heated models and directions of the photon packages created at the
model outer boundary. For radial discretisation $r_i$, the relative projected
area of the innermost cells is much smaller ($\sim r_i^2$), which leads to
a rapid increase in the noise (similar to seen in Fig.~\ref{fig:timeit_bg}).
Therefore, to reduce the maximum temperature errors, one should send more
photon packages that reach the model centre. This works better at low optical
depths, when the initial directions of packages are not randomised by
scattering. When similar weighting is implemented in DIES, the temperature
errors can indeed be decreased almost by a factor of two for a given number
of simulated photon packages (for $k_{\rm bg}=0.2$,
Fig.~\ref{fig:plot_error}a). This is not yet a very significant improvement
for the examined cloud models (Fig.~\ref{fig:plot_error}a) but could be
relevant for models with lower optical depths and larger ratios between the
sizes of the smallest cells and the entire model.

In \citet{Juvela2005}, the most important weighting scheme for TMC calculations
was the location of re-emitted photons. This was crucial when a point source
was surrounded by a thin layer of hot dust. Without weighting, the probability
of a photon package being created in this layer is proportional to cell volume
($\sim r_i^{-3}$ compared to $r_i^{-2}$ for hits by external photons).
However, this weighting is only applicable to the TMC scheme, where the
positions of the re-emitted photon packages can be created independently. In
IRE, re-emission takes place at the location of the absorption events and is
directly dictated by the preceding path of the photon packages. 
However, this also makes the IRE scheme effective in sampling re-emission
from hot dust, which might cover only a small fraction of the whole model
volume.

The other options of weighted sampling (e.g. direction of emitted or scattered
photon packages) were not found to be useful in our tests. In particular, in the
case of scattering, the mismatch between the enforced directional distribution
and the dust scattering function (usually with $g\gg 0$) results in a small
number of photon packages with large weights, with negative effects (shot
noise). 

There may also be more general reasons why weighted sampling would be less
effective in IRE calculations. One potential factor is the way a photon
package is assigned a frequency. In the traditional Monte Carlo, each package
has a fixed frequency, and only the weight of the photon package changes as it
moves through the model. In contrast, when an IRE package encounters an
absorption event, the result is re-emission at a random frequency, which is
dictated by the change in the cell emission spectrum (which results from the
temperature rise in the cell).  One is thus not likely to simulate many
consecutive short steps. After the first absorption, re-emission is already at
a longer wavelength and with a longer free path. The opposite is also possible.
Due to the low optical depth at long wavelengths, a long-wavelength
photon package can propagate relatively freely deep into the model. With some
probability, it can at re-emission be transformed to a higher frequency, thus
also contributing to that part of the local radiation field.

The basic IRE method works well in optically thick models. On the other hand,
at low optical depths, the photon packages only
rarely interact with the medium, and it can be beneficial to increase the number of interaction events
by biasing. The modification of the probability distribution of free paths
with a constant factor $k_{\rm A}$ (for modified probability $p=e^{-k_{\rm A}
\tau}$ of free paths) did not perform particularly well, especially compared
to a variable and density-dependent modification with $k_{\rm A}(\rho)$. In
other words, it seems important that weighting is used only in those parts of
the model where it is needed. The density-dependent weighting with $k_{\rm
A}(\rho)$ still requires that we set parameters, how much weighting is
applied, and how that depends on the density. Although the results were good in
the test cases, the same parameter values might not work well in other cases
and might need to be tuned separately. Therefore, FFI appears to be a good
alternative; it performed almost as well as the density-dependent weighting
for optically thin models, and even slightly better for the optically thick
models, and has no free parameters. The method requires some additional
computations, but the overheads are barely visible in our tests.

There are still other techniques that might improve the efficiency of IRE
calculations, at least for certain types of models. One such method is photon
splitting, which could enhance the radiation field sampling in selected parts
of the model. However, this would be slightly more difficult to implement
efficiently, especially on GPUs, where the goal is to have all the threads
always performing strictly the same operations. We also did not consider the
frequency distribution of the re-emitted photon packages. Weighted frequency
sampling may not improve the accuracy of temperature estimates but, for
example, could be helpful to produce better estimates of scattered light in
given frequency bands. However, due to its use of a fixed grid of frequencies,
the TMC method appears better suited for the generation of images of scattered
light.

Although we have  examined only spherically symmetric models in this paper, all
the discussed methods can in principle be used in three-dimensional RT calculations. In any model, it is important to consider which
directions or regions should be emphasised with the weighted sampling.

Finally, we note that there is one small difference between TMC and IRE that could lead to differences in their results. In the IRE method, re-emission takes place
at the locations where photon packages are absorbed, whereas in TMC this happens uniformly over
the cell. This affects the fraction of re-emission that is subsequently reabsorbed in
the same cell or transported to the next cells. In this respect, the IRE way is
more correct; although, since each cell is still assigned only a single
temperature, neither method can actually resolve differences at subcell
scales. We tested this by modifying IRE so that the position of re-emission
was randomised within the cell volume. No visible differences were observed
for any of the models examined in this paper. However, with sufficiently high optical depths and sufficiently coarse discretisation, the difference could become
noticeable, although the actual errors would probably still be dominated
directly by the poor discretisation.

\section{Conclusions} \label{sect:conclusions}

We present DIES, a parallel implementation of the IRE method for dust
RT calculations. Here we use the program to study the
potential benefits of weighted sampling. This work led to the following
conclusions:
\begin{itemize}
\item  IRE is also well suited for GPU computing. For one-dimensional models of
moderate size ($\sim 100$ shells), the RT calculations could
be completed within a few seconds, with temperature uncertainties being sufficiently low
for most applications ($\delta T \la 0.1$\,K).
\item Weighted sampling is usually not needed for models of high or moderate
optical depth. In particular, in case of embedded point sources, the IRE
method itself already provides efficient sampling of the dust re-emission, even
when this is confined to a very small volume.
\item There can be  some advantage from biasing the directions of the 
photon packages representing the external radiation field. This is true
especially for optically thin models and when the projected area of the
innermost model cells is very small.
\item The IRE method samples absorptions only at discrete positions, which can
result in larger temperature errors in optically thin models. The FFI
technique (i.e. forcing at least one interaction for each photon package) is
a simple way to reduce the noise. Density-dependent biasing of the photon free
paths can be equally beneficial but requires some tuning of the method
parameters.
\end{itemize}

\begin{acknowledgements}

MJ acknowledges the support of the Research Council of Finland Grant No. 348342.

\end{acknowledgements}

\bibliography{my.bib}

\begin{appendix}

\section{Comparison to CRT program}  \label{app}

We compared the run times of DIES to those of the CRT program
\citep{Juvela2005}.  The chosen model clouds is a 1\,$M_{\odot}$ Bonnor-Ebert
sphere that has been divided into 100 concentric shells of equal thickness.
The model is illuminated either only by an external isotropic radiation field
or by a central point source, as described in Sect.~\ref{sect:results}. The
runs were performed on the same laptop, CRT runs using the CPU and DIES the
laptop GPU.

\begin{figure}
\begin{center}
\includegraphics[width=9.0cm]{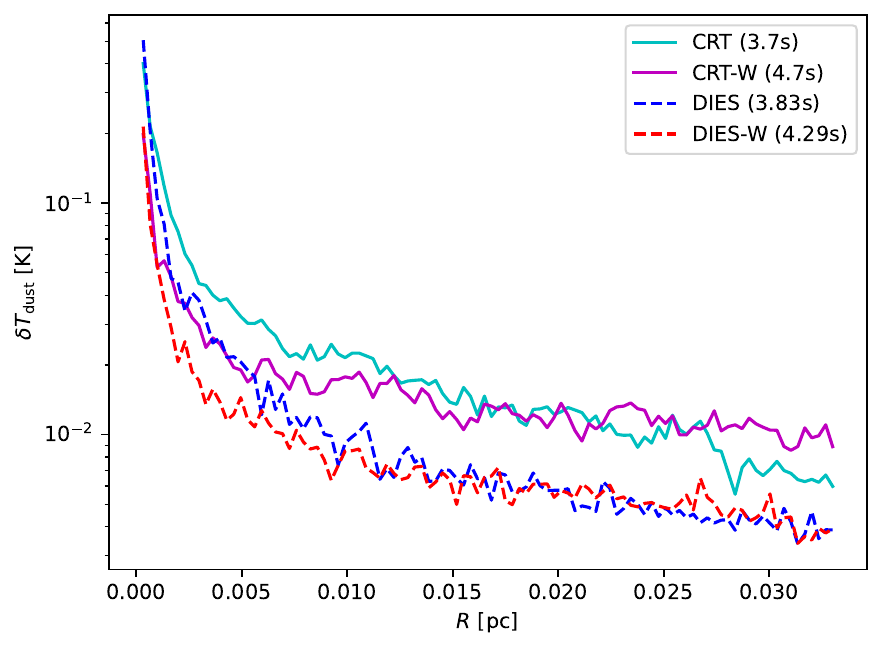}
\end{center}
\caption { 
Noise of dust temperature estimates, $\delta T$, as a function of the radius
in an externally heated 1\,$M_{\odot}$ model cloud. Results are shown for the
CRT program (solid lines) and the DIES program (dashed lines), with the run
times quoted in the legend. Results are shown also for weighted runs, when
photon packages are targeted more towards the model centre (``CRT-W'' and
``DIES-W'', respectively). The noise has been estimated from the standard
deviation between 50 runs.
}
\label{fig:cmp_bg}
\end{figure}

Figure~\ref{fig:cmp_bg} shows the results for the externally heated model. The
dust temperatures are low, re-emission is insignificant, and therefore CRT
runs require only one iteration. Plot also shows results for alternative runs
with weighted sampling, where photon packages were targeted preferentially
towards the model centre. This has a small but noticeable effect in decreasing
the noise in the innermost shells. The number of photon packages was selected
so that the run times (CRT on CPU and DIES on GPU) were approximately similar.
The resulting noise of the DIES estimates is more than a factor of two lower
in the outer part of the model, thus in theory corresponding to a factor of
four saving in the run time. However, in the centre the noise levels are
identical. We suspect that CRT benefits here from the fact that all
frequencies are sampled equally with the same number of photon packages. In
contrast, DIES creates photon packages according to the spectrum of the
background radiation. This gives more emphasis for the frequencies near the
peak of the emission spectrum, while dust in the model centre is heated mainly
by direct background radiation at somewhat longer wavelengths. This suggest
that one should consider weighting the sampling also as a function of the
frequency, especially in case of optically very thick models. As mentioned
above, re-emission plays no role in the case of the externally heated model.

\begin{figure}
\begin{center}
\includegraphics[width=9.0cm]{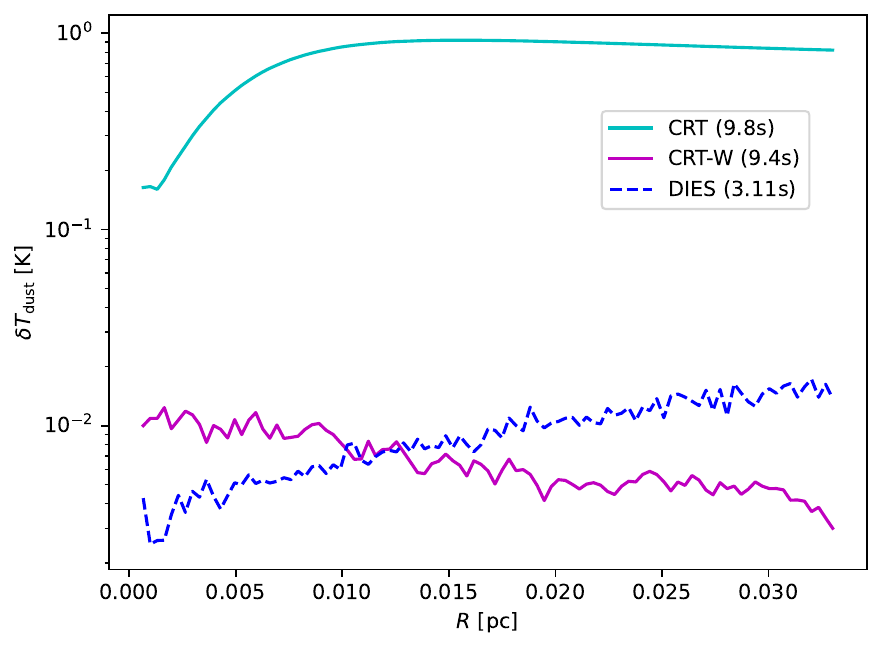}
\end{center}
\caption { 
As Fig.~\ref{fig:cmp_bg} but for a model with a central radiation source and
no external radiation field.
}
\label{fig:cmp_ps}
\end{figure}

Figure~\ref{fig:cmp_ps} shows a similar comparison when the model cloud is
heated only by a central radiation source.  The basic IRE method already
samples the re-emission well, and there is no need for weighted sampling.  In
contrast, CRT shows in the unweighted case large noise and very strong bias.
This is because, when the re-emission is sampled uniformly over the model
volume, only few photon packages are created in the hot shells near the
central radiation source. However, weighted sampling reduces the CRT noise by
two orders of magnitude. In this case the weighting means that the same number
of re-emitted photon packages were created for each shell, irrespective of
their volumes \citep[cf.][]{Juvela2005}. The quoted CRT run times are for two
iterations. The first iteration gives initial estimates for the dust
temperatures, and only the second iteration gives the first indication of how
these are affected by re-emission. However, several more iterations would be
needed to reach final convergence in this test case, and the real cost of CRT
calculations would be a few times higher than indicated in the figure. This
remains true, even though in multi-iteration runs the number of photon
packages per iteration could be decreased with the help of a reference field,
as discussed in \citet{Juvela2005}. Thus, the IRE method is simpler and likely
to be much more efficient when the model includes strong local heating.
However, the methods are also seen to prioritise different parts of the model.
IRE is more accurate in the inner regions, while the standard Monte Carlo
simulation could still in some cases be superior in the outer parts.

\end{appendix}

\end{document}